\documentstyle[camera,prbplug,psfig]{article}
\def\csvo{Ca$_{1-x}$Sr$_{x}$VO$_{3}$}
\def\cvo{CaVO$_3$}
\def\svo{SrVO$_3$}
\def\slto{Sr$_{0.95}$La$_{0.05}$TiO$_{3}$}
\def\yto{YTiO$_{3}$}
\def\neff{$N_{\rm eff}$}
\def\meff{$m^{\ast}$}
\begin{document}
\draft
\title{
Band-width control in a perovskite-type 3$d^{1}$ correlated metal 
Ca$_{1-x}$Sr$_{x}$VO$_{3}$. II.\\ Optical spectroscopy investigation.
}
\author{H.~Makino}
\address{
Institute of Applied Physics, University of Tsukuba, Tsukuba 305, Japan\\
}
\author{I.~H.~Inoue\cite{a.etl}~\cite{a.inoue}}
\address{
PRESTO Japan Science and Technology Corporation, 
and Electrotechnical Laboratory, Tsukuba 305, Japan\\
}
\author{M.~J.~Rozenberg}
\address{
Institut Laue Langevin, B.P. 156X, 38042 Grenoble C\'{e}dex 9, France\\
}
\author{I.~Hase\cite{a.etl} and Y.~Aiura\cite{a.etl}}
\address{
Electrotechnical Laboratory, Tsukuba 305, Japan\\
}
\author{S.~Onari}
\address{
Institute of Applied Physics, University of Tsukuba, Tsukuba 305, Japan\\
}
\date{December 25, 1997}
\maketitle
\begin{abstract}
Optical conductivity spectra of single crystals of the 
perovskite-type 3$d^{1}$ metallic alloy system \csvo\ have been studied to 
elucidate how the electronic behavior depends on the strength of the 
electron correlation without changing the nominal number of electrons.
The reflectivity measurements were made at room temperature between 
0.05~eV and 40~eV.
The effective mass deduced by the analysis of the 
Drude-like contribution to the optical conductivity and 
the plasma frequency do not show critical enhancement, 
even though the system is close to the Mott transition.
Besides the Drude-like contribution, two anomalous features 
were observed in the optical conductivity spectra of 
the intraband transition within the 3$d$ band.
These features can be assigned to transitions involving the incoherent 
and coherent bands near the Fermi level.
The large spectral weight redistribution in this system, however, 
does not involve a large mass enhancement.
\end{abstract}
\section{INTRODUCTION}
Over the past few decades, a considerable number of studies 
have been performed on 3$d$ transition-metal oxides which have a 
considerably narrow 3$d$ band.
In particular, a metal-to-insulator transition 
caused by a strong electron correlation\cite{mott} (Mott transition) 
as well as anomalous electronic properties in the metallic phase near 
the Mott transition have attracted the interest of many researchers.
Since the discovery of the high-$T_{c}$ cuprate 
superconductors, the importance of two types of experimental 
approaches to the Mott transition have been discussed intensively:\ 
a filling control and a band-width control. 
The former consists in doping holes or electrons to 
the system, and the latter in varying 
the strength of the electron correlation $U/W$, 
where $U$ is the electron correlation due to Coulomb repulsion 
and $W$ is the one-electron band-width.

In recent years, the systematic evolutions of optical conductivity spectra 
in going from a correlated metallic phase to the Mott-Hubbard 
insulating phase have been reported on both the filling 
controlled\cite{fujishima1992,taguchi1993,kasuya1993,crandles1994,%
katsufuji1995} 
and the band-width controlled Mott-Hubbard 
systems.\cite{crandles1991,okimoto1995,thomas1994,rozenberg1995}
The smallest energy gap for charge excitations of the Mott-Hubbard 
insulator is the excitation energy 
of the charge fluctuation $d^n+d^n \rightarrow d^{n-1}+d^{n+1}$, 
so-called a Mott-Hubbard gap.\cite{zsa1985}
The optical conductivity of the Mott-Hubbard insulator is 
considered to show a gap feature due to the above charge excitation 
from the lower Hubbard band to the upper Hubbard band.

V$_{2}$O$_{3}$ and related compounds have been extensively studied 
as typical materials which show the Mott transition with varying 
the strength of the $U/W$ ratio.
The temperature dependent optical conductivity of V$_{2}$O$_{3}$ 
was reported by Thomas {\it et al\/}.\ through the metal-to-insulator 
transition.\cite{thomas1994,rozenberg1995}
The optical conductivity of V$_{2}$O$_{3}$ in the insulating phase shows 
a gap feature, which is attributed to the Mott-Hubbard gap excitation. 
On the other hand, in the metallic state, a low-energy contribution 
to the optical conductivity shows an anomalous feature, which 
is reproduced by two Lorentzians.
Moreover, the optical conductivity shows an anomalous enhancement 
of the spectral weight as a function of temperature.
Rozenberg {\it et al\/}.\ reported that these experimental results 
are in good agreement with the theoretical prediction 
obtained by the infinite-dimension Hubbard model within the 
mean-field approach.\cite{rozenberg1995}
Since the formation of the Mott-Hubbard bands are predicted even 
in the metallic state for this kind of strongly correlated system, 
the optical conductivity spectra 
should be affected by the precursor features.\cite{georges1996}
It is very interesting to see how the spectral weight varies with the 
electron correlation in the correlated metallic state near the 
Mott transition.
For a detailed discussion, however, we need to control the strength of 
the electron correlation more precisely.

The perovskite-type early 3$d$ transition-metal oxides are ideal 
Mott-Hubbard systems for controlling the band filling and band-width 
by chemical substitutions.
It has been reported that, for the filling controlled systems 
La$_{1-x}$Sr$_{x}$TiO$_{3}$ and 
$R_{1-x}$Ca$_{x}$TiO$_{3}$ ($R$=rare earth), 
the spectral weight of the optical conductivity 
transfers from the higher energy feature corresponding to an 
excitation through the Mott-Hubbard gap, 
to the mid-infrared inner-gap region corresponding to 
the Drude-like absorption extending 
from $\omega=0$.\cite{fujishima1992,taguchi1993,katsufuji1995}
The rate of the spectral weight transfer by doping   
increases systematically with the increase of the one-electron 
band-width $W$.

On the other hand, the Ti--O--Ti bond angle can be decreased 
as we decrease the ionic radius of the $R$ site.
The decrease of the Ti--O--Ti bond angle gives rise to 
the decrease of the value of $W$.
A systematic change of the optical conductivity spectra 
was reported on  
$R$TiO$_{3}$ ($R$=La, Ce, Pr, Nd, Sm, and Gd).\cite{crandles1991}
The lowest gap-like feature systematically increases as the ionic radius 
of the $R$ site decreases, namely, as the value of $W$ decreases.
A similar change was also observed on 
an alloy system, La$_{1-x}$Y$_{x}$TiO$_{3}$.\cite{okimoto1995}
These systematic variations of the optical conductivity are interpreted 
as the successive increase of the Mott-Hubbard gap with the increase 
of the strength of the $U/W$ ratio.
In these materials, however the system remains an insulator even 
for the least correlated LaTiO$_{3}$, 
therefore one cannot study the evolution of the metallic properties 
under the band-width control in this system.

The purpose of this paper is to clarify the evolution of the optical 
conductivity spectrum in the metallic phase near the Mott transition, 
as we control the strength of the electron correlation $U/W$ without 
changing the band filling.
A perovskite-type $3d^{1}$ vanadate \cvo\ is considered to be 
a strongly correlated metal close to the Mott 
transition.\cite{inoue1993,inoue0000}
We can control the strength of the $U/W$ ratio,
by chemical substitution of a Sr ion for a Ca ion of the same valence
without varying the nominal $3d$-electron number
per vanadium ion.\cite{inoue1995}
We report the optical conductivity spectra 
in this strongly correlated alloy system \csvo.
The effective mass \meff\ estimated from the optical measurements
are shown in Sec.~\ref{subsec:mass}\@. 
The evolution of the optical conductivity is discussed in 
Sec.~\ref{subsec:spectral}\@. 
\section{Experimental}
\label{sec:experimental}
Single crystals of \csvo\ ($x$=0, 0.25, 0.5, 1) were 
grown by a floating-zone method using  
an infrared image furnace with double halogen lamps.
Details on the preparation of these samples are described 
in the preceding paper.\cite{inoue0000}
Since as-grown samples are slightly oxygen deficient,
all the samples were annealed in air at 200~$^{\circ}$C 
for 24~hours in order to make the oxygen concentration
stoichiometric.\cite{inoue1993,fukushima1994,shirakawa1995}

Raman scattering spectra were measured at room temperature 
in back-scattering geometry 
using a triple spectrometer system (Jasco TRS-600) 
equipped with a charge coupled device (CCD; Photometrics TK512CB) 
cooled by liquid nitrogen. 
The samples were excited by the 514.5~nm Ar ion laser line. 
Polarization of the incident light was taken to be parallel
to that of the scattered light.

Optical reflectivity measurements were carried out 
at room temperature ($\sim$\,300~K)
in the energy range between 0.05~eV and 40~eV 
using a Michelson-type Fourier-transform 
infrared spectrometer (0.05--0.6~eV),
a grating monochrometer (0.5--5.6~eV), 
and a Seya-Namioka-type grating for the synchrotron 
radiation (5--40~eV) at the beamline BL-11D of Photon
Factory, Tsukuba. 
The surfaces of the samples were mechanically polished
with diamond paste for the optical measurements.
The absolute reflectivity was determined by referring  
to the reflectivity of an Al or Ag film 
which was measured at same optical alignment.

We have calculated a complex dielectric function 
$\epsilon(\omega)\equiv\epsilon_{1}(\omega)+i\epsilon_{2}(\omega)$ 
by the Kramers-Kronig (K-K) transformation of  
the measured reflectivity $R(\omega)$, where $\omega$ is the photon 
energy.
The real part of the complex optical conductivity 
Re$[\tilde{\sigma}(\omega)]$ is related to 
the imaginary part of the dielectric function $\epsilon_{2}(\omega)$ 
by Re$[\tilde{\sigma}(\omega)]= (\omega/4\pi)\epsilon_{2}(\omega)$.
Since the K-K analysis requires $R(\omega)$ for $0<\omega<\infty$, 
two assumptions must be made to extrapolate the observed data 
beyond the upper and lower bounds of the measurements.
In the present study, the reflectivity data were first extrapolated from 
the lowest measured energy down to $\omega=0$ with the Hagen-Rubens 
formula which is an approximation for $R(\omega)$ of conventional 
metals. 
Then, beyond the highest measured energy, the reflectivity data were 
extrapolated up to $\omega \rightarrow \infty$ with an 
asymptotic function of $\omega^{-4}$.
\section{RESULTS AND DISCUSSION}
\label{sec:results}
\subsection{Band-width control due to orthorhombic distortion}
\label{subsec:bandwidth}
\begin{figure}
    \centerline{\psfig{file=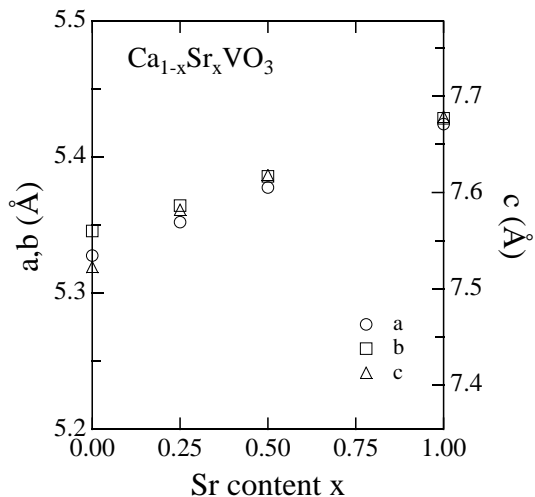,width=7cm}}
    \caption{Lattice constants, $a$, $b$, and $c$ of the  
    Ca$_{1-x}$Sr$_{x}$VO$_{3}$ single crystals plotted against 
    the Sr content $x$. The data were estimated from powder x-ray 
    diffraction patterns.}
    \label{f.Lattice}
\end{figure}

Powder x-ray diffraction measurements were carried out 
to characterize the samples and to determine the lattice constants.
In Fig.~\ref{f.Lattice},  
we present the lattice constants, $a$, $b$, and $c$ 
against the Sr content $x$.
The crystal structure of \csvo\ belongs to  
the perovskite-type structure with orthorhombic 
distortion (GdFeO$_{3}$-type).\cite{chamberland1971}
The amount of the distortion is almost proportional to the amount of 
the Ca content; {\it i.e.}, \svo\ is a cubic perovskite.
However, we assumed 
the crystal structure of all samples ($0\leq x\leq 1$) 
orthorhombic and deduced the lattice constants.
The lattice constants increase monotonously  
with the increase of $x$, ensuring the appropriate  
formation of the solid solution over the whole composition range.

\begin{figure}
    \centerline{\psfig{file=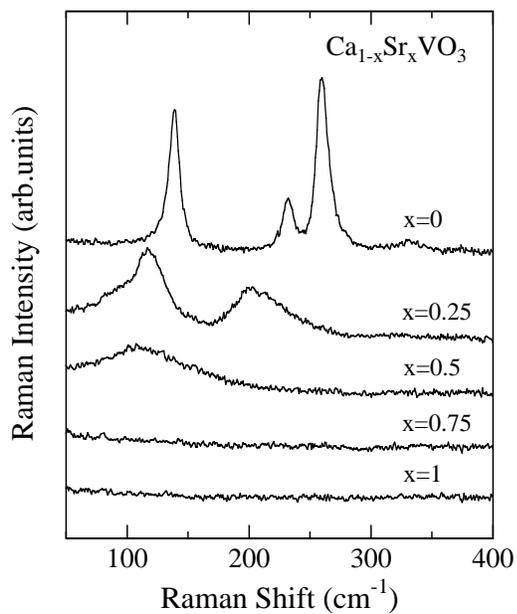,width=7cm}}
    \caption{Raman spectra of Ca$_{1-x}$Sr$_{x}$VO$_{3}$ 
    at room temperature. Polarization of the incident light was 
    taken to be parallel to that of the scattered light.}
    \label{f.Raman}
\end{figure}

Raman scattering measurements give indirect information 
about the crystal symmetry, because the appearance of some Raman-active 
phonon lines depends crucially on the crystal symmetry of the system.
In Fig.~\ref{f.Raman}, we show the Raman spectra of \csvo\ 
at room temperature in the wave-number range of 80--400cm$^{-1}$. 
In this wave-number range, four Raman active $A_{g}$ phonon lines 
are observed in orthorhombic \cvo\ ($x=0$).
The energies of the phonon lines shift to the lower energy side, and 
the width of the peaks become broader, 
as we increase the Sr content $x$.
The Raman active phonon lines disappear completely in \svo.

According to the group theory analysis,\cite{couzi1972} it is predicted that 
several Raman active phonon modes
$7A_{g}+7B_{1g}+5B_{2g}+5B_{3g}$
can exist in the orthorhombically distorted perovskite
which belongs to the point group symmetry of $D_{2h}$\@.
On the other hand, the cubic perovskite which belongs to 
the point group symmetry of $O_{h}$ is ``Raman forbidden'', 
namely it has no Raman-active phonon modes.
Therefore, the experimental results tell us that the crystal symmetry 
actually changes 
from the orthorhombic distorted perovskite (\cvo) 
to the cubic perovskite (\svo)\@. 
The orthorhombic to cubic transition is considered to 
occur between $x=0.5$ and 0.75.

As discussed in the preceding paper,\cite{inoue0000} 
the orthorhombic distortion implies that 
the V--O--V bond angle is deviated from 180$^\circ$, {\it i.e.}, there is 
an alternately tilting network of the VO$_{6}$ octahedra.
The V--O--V bond angle of \svo\ is 180$^{\circ}$ same as an ideal 
perovskite structure, while that of \cvo\ is $\sim$\,160$^{\circ}$.
The bond angle deviation from 180$^\circ$ reduces 
the overlap between 
the neighboring V~3$d$ orbital mediated by the O~2$p$ orbital.
Therefore, the one-electron band-width $W$ of V 3$d$ band decreases 
with  decreasing the V--O--V bond angle.
Accordingly, we can control the value of $W$ 
by chemical substitution of the Ca$^{2+}$ ion for the Sr$^{2+}$ ion 
of the same valence without varying the nominal $3d$-electron number 
per vanadium ion.
Since the electron correlation energy $U$ are almost the same  
in \cvo\ and \svo, we can thus control the strength of 
the $U/W$ ratio by the chemical substitution.
The V--O--V bond angle of \cvo, in addition, is almost equal to that of 
LaTiO$_{3}$ which is a Mott-Hubbard type insulator, so that 
it seems reasonable to consider that \cvo\ is close to the Mott transition.
Moreover, there are many other evidences of the strong electron 
correlation in this system discussed so 
far.\cite{inoue1993,inoue0000,inoue1995,fukushima1994,shirakawa1995}
Thus, \csvo\ system is ideal for the study of the metallic states 
near the Mott transition.
%
\subsection{Effective mass}
\label{subsec:mass}
\begin{figure}
    \centerline{\psfig{file=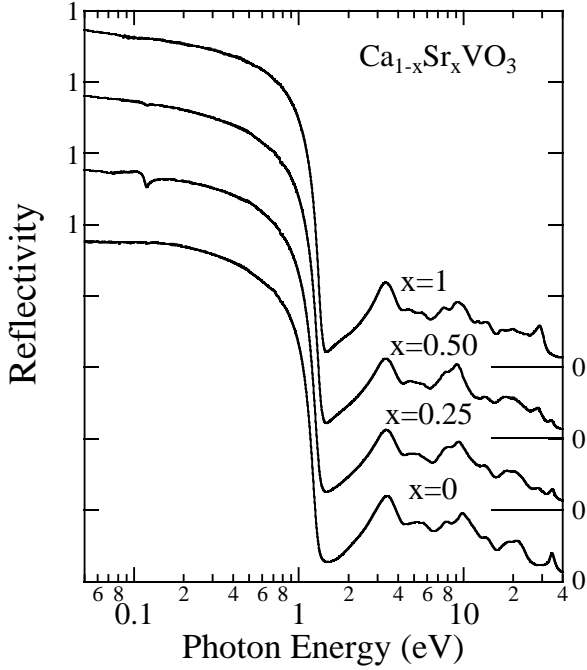,width=8cm}}
    \caption{Reflectivity spectra for the 
    Ca$_{1-x}$Sr$_{x}$VO$_{3}$ single crystals measured 
    at room temperature. The feature at $\sim$\,0.1~eV for $x=0.25$ 
    is an experimental artifact.}
    \label{f.Reflection}
\end{figure}

In Fig.~\ref{f.Reflection}, we show optical reflectivity spectra of \csvo\ 
measured at room temperature ($\sim$\,300~K), 
for four single crystals with different $x$ ($x$=0, 0.25, 0.50, 1)\@.
The chemical substitution of Sr$^{2+}$ for Ca$^{2+}$ seems to make no 
remarkable change at the lower-energy region (below $\sim$\,5~eV) 
in the optical reflectivity spectra.
All the samples exhibit high reflectivity 
from far-infrared to near-infrared region,
and we can recognize a sharp reflectivity edge
appearing at $\sim$\,1.3~eV\@.
Since \csvo\ is metallic over 
the whole composition range, the optical reflectivity 
is dominated by the signal of conduction electrons 
in this photon energy region.
Systematic spectral changes with $x$ are observed 
in the energy range of the ultra-violet and vacuum-ultra-violet light.
The changes are partly due to the differences 
in the conduction bands of the Ca$^{2+}$ and Sr$^{2+}$ cations.
But this is irrelevant for the discussion of the main subject.

First of all, we will concentrate on the low energy response of 
the itinerant carriers.
The contribution of the conduction electrons to the complex dielectric 
function $\epsilon(\omega)$ is well described by the Drude model.
According to the generalized Drude model,\cite{allen1977,webb1986} 
$\epsilon(\omega)$ is expressed as 
\begin{equation}
\epsilon(\omega)=\epsilon_{\infty}
-\frac{4\pi \tilde{\sigma}(\omega)}{i\omega}\equiv \epsilon_{\infty}
-\frac{\omega_{p}^{2}(\omega)}{i\omega(\gamma(\omega)-i\omega)},
\label{epsilon}
\end{equation}
where $\epsilon_{\infty}$ is the high-energy dielectric constant, 
which is a high-energy contribution of the interband transitions, 
$\tilde{\sigma}(\omega)$ is the complex conductivity, 
$\gamma(\omega)$ is the energy-dependent scattering rate, 
and $\omega_{p}(\omega)$ is the plasma frequency.
The plasma frequency $\omega_p(\omega)$ is defined as
\[\omega_{p}^{2}(\omega)\equiv\frac{4{\pi}ne^{2}}{m^{\ast}(\omega)},\]
where $n$ is the total density of conduction electrons, 
and $m^{\ast}(\omega)$ is the energy-dependent effective mass.

To begin, let us confine our attention to 
the plasma frequency.
If we assume that the nominal electron number per vanadium ion is 
exactly 1 for the whole composition range, 
we can deduce the carrier density $n$ from the unit-cell volume.
Then, we can estimate a variation of the effective mass 
$m^{\ast}(\omega)$ from the value of $\omega_{p}(\omega)$.

The Energy-loss function Im($-1/\epsilon$) is obtained 
by the Kramers-Kronig analysis of the measured reflectivity spectra 
$R(\omega)$\@.
Provided that $\gamma(\omega)$ and $m^{\ast}(\omega)$ do not 
depend on $\omega$ strongly, 
we can estimate $\omega_p (=const.)$ from the Energy-loss function, 
because, the Energy-loss function peaks at the energy of 
$\omega_{p}^{\ast}$ 
($\omega_{p}^{\ast}=\omega_p/\sqrt{\epsilon_{\infty}}$).
Accordingly, we can obtain the energy-independent plasma frequency 
$\omega_{p}$ from the peak position of the energy-loss function.

\begin{figure}
    \centerline{\psfig{file=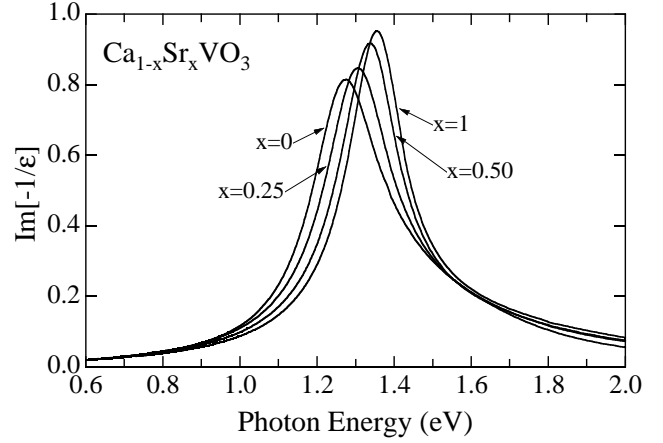,width=\columnwidth}} 
    \caption{Energy-loss function Im[-1/$\epsilon(\omega)$]
    obtained by the Kramers-Kronig transformation of the 
    reflectivity data. The data in the photon energy range 
    between 0.6~eV and 2.0~eV are shown to focus on 
    the plasmon peak (around 1.3~eV).}
    \label{f.Im1e}
\end{figure}

In Fig.~\ref{f.Im1e}, the spectra of Im($-1/\epsilon$) of \csvo\ are 
shown in the photon energy range from 0.6 to 2.0~eV to focus on 
the peak near the reflectivity edge (around 1.3~eV)\@.
The peak position of the Energy-loss function $\omega_{p}^{\ast}$ 
systematically shifts to higher energy with increasing $x$.
At first, we have estimated $\omega_{p}^{\ast}$ from the peak energy, 
and deduced $\omega_{p}=\sqrt{\epsilon_{\infty}}\omega_{p}^{\ast}$.
If we consider only the response of the conduction electrons, 
$\epsilon_{\infty}$ is the contribution from the high-energy interband 
transitions.
Since the interband transition appeared above $\sim\,$2.5~eV, 
the value of $\epsilon_{\infty}$ can be taken from the real part of 
the dielectric function $\epsilon_{1}(\omega)$ 
at around 2.5~eV.\cite{epsilon}
Since the value of $\epsilon_{1}(\omega \sim 2.5 \rm{eV})$ 
is nearly independent of $x$\@, we have used $\epsilon_{\infty}=4$ 
over the whole composition range.
Then, we can deduce the value of \meff\ using the lattice 
constants and the value of $\omega_{p}$.

\begin{figure}
    \centerline{\psfig{file=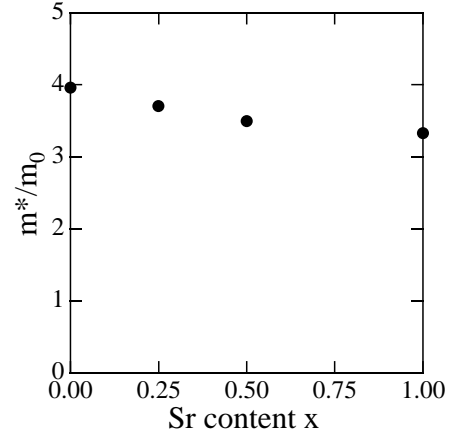,width=6cm}}
    \caption{Effective mass $m^{\ast}$ estimated by the plasma 
    frequencies compared with the bare electron mass $m_{0}$. 
    The value of $m^{\ast}/m_0$ 
    systematically increases in going from SrVO$_{3}$ ($x=1$) 
    to CaVO$_{3}$ ($x=0$).}
    \label{f.MassWp}
\end{figure}

In Fig.~\ref{f.MassWp}, the ratios of the deduced effective mass 
$m^{\ast}$ to the bare electron mass $m_{0}$ 
are plotted as a function of the Sr content $x$.
The value of $m^{\ast}/m_{0}$ systematically 
increases as varying $x$ from \svo\ to \cvo.
This carrier-mass enhancement, however, is not so large, even though 
the system is near the Mott transition.
This result is consistent with the value of \meff\ estimated 
from the results of the specific heat measurements.\cite{inoue0000}

\begin{figure}
    \centerline{\psfig{file=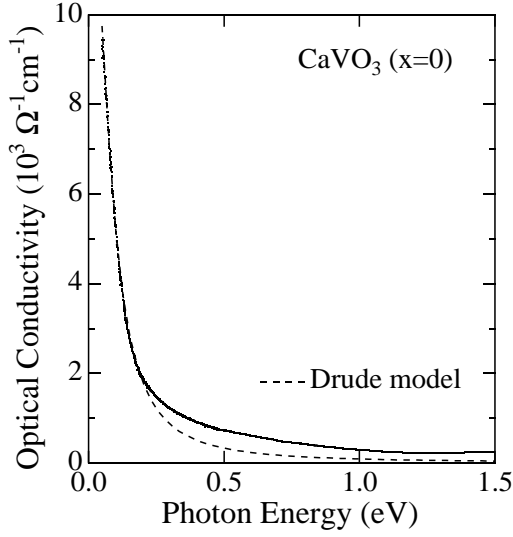,width=7cm}}
    \caption{Comparison between the optical conductivity calculated 
    by the simple Drude model 
    and that of the experiment of CaVO$_{3}$.}
    \label{f.nonDrude}
\end{figure}

It is instructive to compare the measured low frequency 
$\tilde{\sigma}(\omega)$ with the simple Drude model, 
in which $\gamma(\omega)$ and $m^{\ast}(\omega)$ 
do not depend on $\omega$.
According to Eq.~(\ref{epsilon}), the real part of the optical conductivity 
Re$[\tilde{\sigma}(\omega)] \equiv \sigma(\omega)$ 
is given by the formula 
\[\sigma(\omega)=\frac{\sigma_{dc}}{1+\omega^{2}/\gamma^{2}},\]
where $\sigma_{dc}$ is the dc conductivity.
The dc conductivity is expressed by the scattering rate $\gamma$ 
and the plasma frequency $\omega_{p}$ by the following relation:
\[\sigma_{dc}=\frac{ne^{2}}{m^{\ast}\gamma}=
\frac{\omega_{p}^{2}}{4\pi \gamma}.\]
Here, we have used the value of $\sigma_{dc}$ obtained 
by electric resistivity measurements at room temperature.
Then the value of $\gamma$ can be deduced from the above relation.
Fig.~\ref{f.nonDrude} shows the comparison 
of the experimentally obtained $\sigma(\omega)$ for \cvo\ 
to the optical conductivity calculated by the simple Drude model.
As shown in Fig.~\ref{f.nonDrude}, 
the contribution to the optical conductivity below $\sim\,$1~eV\@, 
which is considered to be a response of the itinerant carriers, 
is not properly reproduced with the simple Drude model 
especially above 0.2~eV\@.
As we increase the photon energy, the experimentally obtained 
$\sigma(\omega)$ deviates from that of the simple Drude model.
The observed $\sigma(\omega)$ has a tail decaying slower than 
the Drude-type $\omega^{-2}$ dependence.

\begin{figure}
    \centerline{\psfig{file=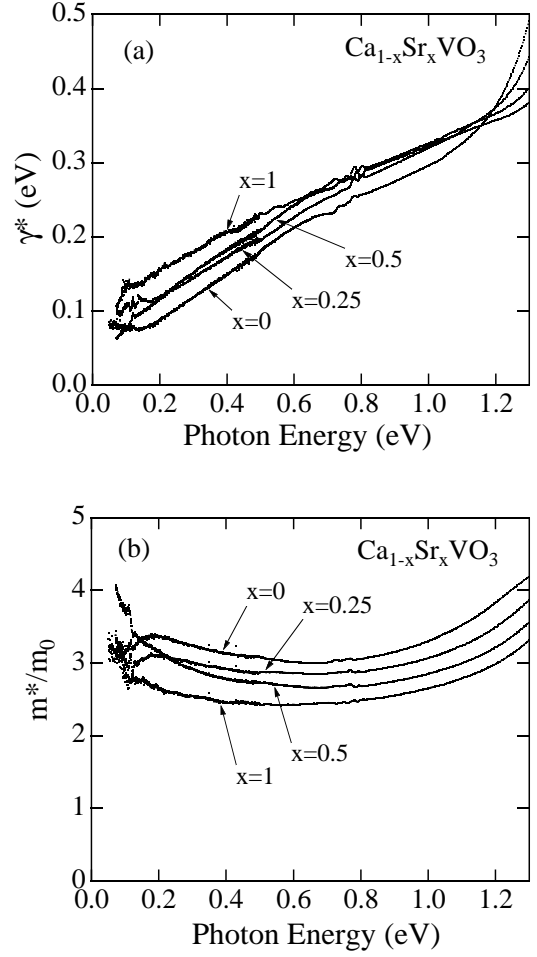,width=7cm}}
    \caption{(a) Energy dependent scattering rates 
    $\gamma(\omega)$ of \csvo; (b) effective mass  
    $m^\ast (\omega)$ of Ca$_{1-x}$Sr$_{x}$VO$_{3}$. 
    $m^\ast (\omega)$ is normalized to the bare electron 
    mass $m_{0}$.}
    \label{f.ExDrudeMG}
\end{figure}

We consider that the discrepancy between the simple Drude model and 
the experimental results is attributed to 
the energy-dependence of the scattering rate $\gamma(\omega)$ 
and the effective mass $m^{\ast}(\omega)$.
The $\gamma(\omega)$ corresponds to the renormalized scattering rate 
$\tau^{\ast-1}(\omega)$\@, which is 
$\tau^{-1}\times(m/m^{\ast}(\omega))$\@.
The quantity $\tau(\omega)$ is closer to the microscopic intrinsic 
quasiparticle lifetime at $T=0$\@.
We can determine $\gamma(\omega)$ and \meff$(\omega)$ 
from Eq.~(\ref{epsilon}); {\it i.e.}, when we define 
$\epsilon(\omega)\equiv\epsilon_{1}(\omega)+i\epsilon_{2}(\omega)$,
\[\gamma(\omega)=\frac{\omega\epsilon_{2}(\omega)}
{\epsilon_\infty-\epsilon_1(\omega)},\]
\begin{eqnarray*}
  m^{\ast}(\omega)&=&\frac{4{\pi}ne^{2}}{\omega^{2}}
  Re\left[\frac{1}{\epsilon_{\infty}-\epsilon(\omega)}\right] \\ %
  &=&\frac{4{\pi}ne^{2}}{\omega^{2}}\,
  \frac{\epsilon_{\infty}-\epsilon_{1}(\omega)}
{(\epsilon_{\infty}-\epsilon_{1}(\omega))^{2}+\epsilon_{2}^{2}(\omega)}.
\end{eqnarray*}
Figs.~\ref{f.ExDrudeMG}(a) and \ref{f.ExDrudeMG}(b) show 
$\gamma(\omega)$ and $m^{\ast}(\omega)$ of \csvo\ 
as a function of photon energy.
In case of the simple Drude model, the scattering rate $\gamma$ is 
provided to be independent on the photon energy.
But, in this system, $\gamma(\omega)$ actually increases as we increase 
the photon energy, as shown in Fig.~\ref{f.ExDrudeMG}(a).

The energy-dependent scattering rate is generally based on the 
electron-phonon scattering and the electron-electron scattering. 
Since an extremely large T$^{2}$-dependence of the dc 
conductivity observed in the \csvo\ system can be well ascribed to 
the electron-electron scattering,\cite{inoue0000}
it is reasonable to consider that the electron-electron scattering 
governs the behavior of $\gamma(\omega)$.
According to the Fermi liquid theory, the electron-electron 
scattering rate is proportional to $\omega^{2}$\@.
Fig.~\ref{f.ExDrudeMG}(a), however, indicates 
that $\gamma(\omega)$ looks more proportional to $\omega$ 
rather than $\omega^{2}$\@.
On the contrary, since the electron-phonon scattering is proportional 
to $\omega^{5}$ up to the Debye frequency, 
it is necessary to elucidate the scattering process which contributes to 
$\gamma(\omega)$.
This is still an open question.

On the other hand, the energy dependence of $m^{\ast}(\omega)$ 
is not so large.
Except for the low energy region ($\omega < \sim0.2$~eV), 
the value of \meff$/m_{0}$ increases with the decrease of Sr content $x$.
In the Sr$_{1-x}$La$_x$TiO$_3$ system, which is a typical doping 
system, as one approaches $x=1$, 
a large energy dependence of \meff\ as well as a critical enhancement 
at the low-energy region are observed.\cite{fujishima1992}
However, in \csvo\ system, \meff\ does not exhibit 
such a critical enhancement with varying $x$ in going from \svo\ to 
\cvo, although there is a difference between the filling control and 
the band-width control.

\begin{figure}
    \centerline{\psfig{file=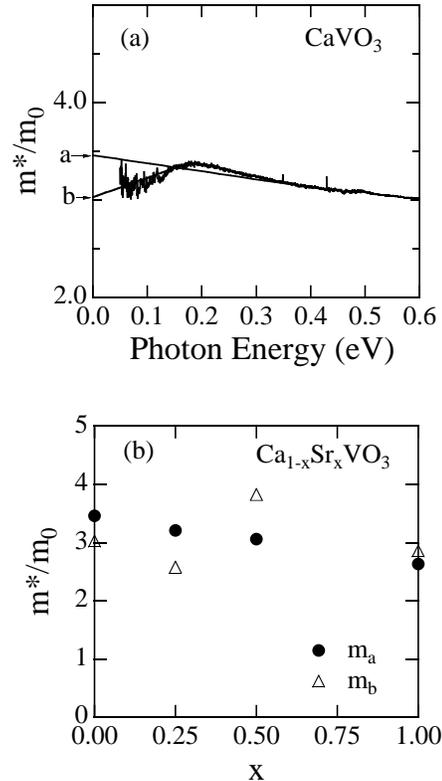,width=6cm}}
    \caption{(a) Energy dependent effective mass $m^{\ast}(\omega)$
    of  CaVO$_{3}$ compared with the bare electron mass. ``a'' and ``b'' 
    indicate intercepts at which tangential lines drawn from 0.4~eV 
    and 0.15~eV cut the vertical axis, corresponding to the values 
    of $m_{a}$ and $m_{b}$.
    (b) $m_{a}$ and $m_{b}$ are plotted against the Sr content $x$.}
    \label{f.MassExDrude}
\end{figure}

In the low energy limit ($\omega=0$), $m^{\ast}$ should correspond to the 
effective mass estimated by the specific heat measurement.
We have interpolated \meff$(\omega)$ down to $\omega=0$ 
with two kinds of tangential lines drawn from 0.4~eV and 0.15~eV\@.
As shown in Fig.~\ref{f.MassExDrude}(a), the intercepts, at which 
the two tangential lines from 0.4~eV and 0.15~eV cut the vertical axis, 
are defined as $m_{a}$ and $m_{b}$\@.
Fig.~\ref{f.MassExDrude}(b) indicates $x$-dependence of 
the values of $m_{a}$ and $m_{b}$\@.
The value of $m_{b}$ does not show a systematic behavior, because phonons, 
randomness, or other extrinsic contributions possibly cause 
this non-systematic change.
However, the value of $m_{a}$ increases systematically with decreasing 
$x$; moreover, the values are almost equal to the value of $m^{\ast}$ 
deduced from plasma frequency.
We regard $m_{a}$ as a good measure of $m^{\ast}/m_{0}$ for this system.

The effective mass estimated from the plasma frequencies and 
the generalized Drude analysis ($m_{a}$) appear in 
Table~\ref{t.Mass}\@.
\begin{table}[b]
  \centering
  \caption{%
         Effective mass $m^{\ast} /m_{0}$ deduced from the plasma 
         frequencies $\omega_{p}$ and the generalized 
         Drude model ($m_{a}$).%
         \label{t.Mass}}
  \begin{tabular}{ccccc}%
   \hline\hline%
    $x$        &     0  &  0.25  &  0.5   &  1    \\ %
   \hline%
    $m^{\ast} /m_{0}$ (deduced from $\omega_{p}$)  
     & 3.9  &  3.7  &  3.5  &  3.3  \\ %
    $m_{a}$ (generalized Drude analysis)  
     & 3.5  &  3.2  &  3.1   &  2.7  \\ %
   \hline\hline%
  \end{tabular}
\end{table}
It is expected that we should observe, near the Mott transition, 
a critical enhancement of the effective mass of the 3$d$ 
conduction electrons.
If we substitute the Ca$^{2+}$ ion for the Sr$^{2+}$ ion in the \csvo\ 
system, the 3$d$ band-width successively decreases.
Then, the value of $m^{\ast} /m_{0}$ is expected to increase 
drastically reflecting the change of the $U/W$ ratio.
But we have observed that such a large mass enhancement 
does not actually take place in this system.
This is consistent with the result shown in the preceding 
paper.\cite{inoue0000}
\subsection{Spectral weight redistribution of 3{\it d\/}-band}
\label{subsec:spectral}
\begin{figure}
    \centerline{\psfig{file=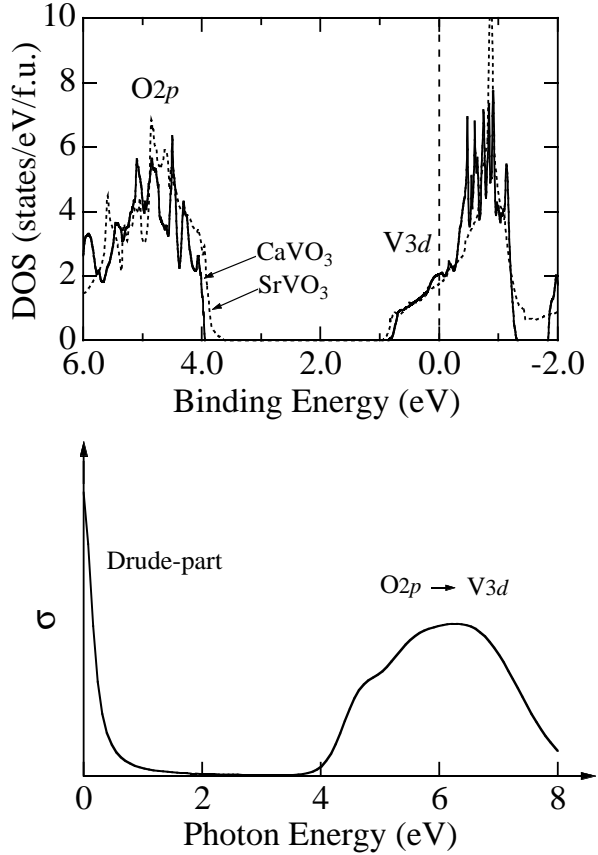,width=8cm}}
    \caption{DOS of CaVO$_{3}$ and SrVO$_{3}$ obtained by 
    the LDA band calculation (top) and a schematic picture 
    of optical conductivity expected from the calculated DOS (bottom).}
    \label{f.DOSandOC}
\end{figure}

The density of states (DOS) of orthorhombic \cvo\ and cubic \svo\ 
calculated using 
the full-potential augmented plane-wave method with the local density 
approximation (LDA) are 
shown in the top of Fig.~\ref{f.DOSandOC}.
The band-calculation shows that the DOS near the Fermi level $E_{F}$ 
is dominated by the V 3$d$ electrons.
The V~3$d$ band crosses the Fermi level, 
and the DOS below 4~eV is mainly the O~2$p$ band.

In the metallic states, $\sigma(\omega)$ is expected to consist of 
two basic components: intraband transitions within the 
V~3$d$ conduction band, {\it i.e.}, the Drude part extending 
from $\omega=0$, 
and interband transitions appearing at much higher energy.
The latter is regarded from the calculated DOS 
as the charge-transfer contribution (an excitation 
from the O~2$p$ band to the unoccupied part of the V~3$d$ band above 
$E_{F}$)\@.
A corresponding schematic picture of the optical conductivity is shown 
in the bottom of Fig.~\ref{f.DOSandOC}.
As seen in the picture, 
the charge-transfer contribution is expected to appear above $\sim$\,4~eV, 
and the absorption edge of the charge-transfer transition in \svo\ 
is considered to shift slightly to lower energy than that of \cvo, 
reflecting the shift of the O~2$p$ band.

\begin{figure}
    \centerline{\psfig{file=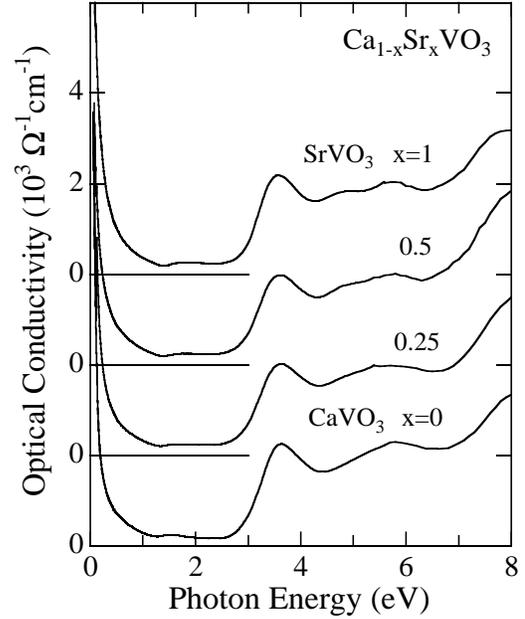,width=7cm}}
    \caption{Optical conductivity spectra of the 
    Ca$_{1-x}$Sr$_{x}$VO$_{3}$ single crystals ($x$=0, 0.25, 0.50, 1) 
    at room temperature obtained by the Kramers-Kronig 
    transformation of the reflectivity data.}
    \label{f.OptiCon}
\end{figure}

Based on this picture, let us now look at the experimental results, 
Fig.~\ref{f.OptiCon} shows the real part of the optical conductivity, 
$\sigma(\omega)$, 
of the \csvo\ single crystals ($x$=0, 0.25, 0.50, 1)\@.
The optical conductivity spectra are different from our naive 
schematic picture (Fig.~\ref{f.DOSandOC} bottom); they show 
the presence of two anomalous features in the intraband transition 
part below 4~eV besides the Drude-like absorption (discussed above): 
a small peak which appears at $\sim$\,1.7~eV and 
a large peak at $\sim$\,3.5~eV.
It must be noted that the two peak-like structures below 4~eV 
have no naive origin as far as we can infer 
from the calculated DOS (Fig.~\ref{f.DOSandOC})\@.
This large spectral weight redistribution is generally believed to be 
a manifestation of the strong electron correlation in this system.

\begin{figure}[b]
    \centerline{\psfig{file=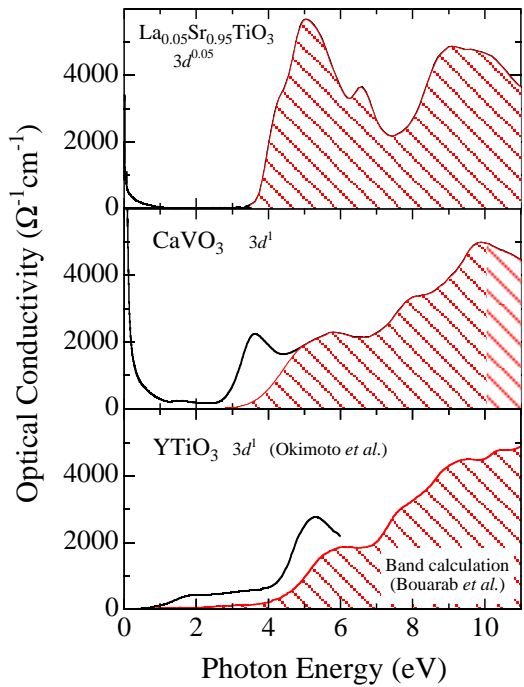,width=7cm}}
    \caption{Comparison of the optical conductivity spectra of 
    CaVO$_{3}$ to those of other perovskite oxides, 
    Sr$_{0.95}$La$_{0.05}$TiO$_{3}$ 
    (lightly doped 3$d^{0}$ metal),\protect\cite{makino0000} 
    and YTiO$_{3}$ (3$d^{1}$ insulator) 
    reported by Okimoto {\it et al\/}.\protect\cite{okimoto1995}
    Shaded portions correspond to the interband transition, and 
    remaining white portions correspond solely to the intraband 
    transitions within the V 3$d$ band.}
    \label{f.CTback}
\end{figure}

Fig.~\ref{f.CTback} shows a comparison of the optical conductivity 
spectra of \cvo\ to those of other perovskite oxides, 
\slto\ (lightly doped 3$d^{0.05}$ metal),\cite{makino0000}
and \yto\ (3$d^{1}$ insulator) reported 
by Okimoto {\it et al\/}.\cite{okimoto1995}
In the optical conductivity of \slto, 
the most prominent low-energy feature, 
that distinctly rises around 4~eV, 
can be interpreted as originating in a transition from the O~2$p$ band to 
the Ti~3$d$ band, which corresponds to the optical gap of the parent 
insulator SrTiO$_{3}$\@.\cite{cardona1965}
The doped 3$d$ electrons contribute to $\sigma(\omega)$ 
with a small spectral weight extending from $\omega=0$.
On the other hand, \yto\ is considered to be a Mott-Hubbard insulator.
Two electronic gap-like features are observed around 1~eV and 4~eV.
These features have been respectively interpreted as originating in 
excitations through the Mott-Hubbard gap, namely, 
from the lower-Hubbard band (LHB) to the upper-Hubbard 
band (UHB)\@, and an excitations through the charge-transfer gap, 
{\it i.e.}, from the O~2$p$ band 
to the UHB.\cite{okimoto1995,arima1993}
Recently, Bouarab {\it et al\/}.\ have reported interband optical 
conductivities obtained by the energy-bands calculation of 
the Y{\it M\/}O$_{3}$ ({\it M\/}=Ti-Cu) 
system with a local spin-density approximation.\cite{bouarab1997}
Their calculated results of interband optical conductivity in \yto\ 
is shown in the bottom of Fig.~\ref{f.CTback} as a shaded portion.
We find out from this comparison that the peak at around 
5~eV cannot be explained by the transition 
between the O~2$p$ band and the Ti~3$d$ band alone.

In \cvo, photoemission spectroscopy\cite{inoue1995,morikawa1995} 
has revealed that the O~2$p$ band is 
located at a binding energy which is almost the same as that of 
metallic \slto; 
hence, the absorption edge of the charge-transfer excitation of \cvo\ should 
be approximately equal to that of \slto\@.
Therefore, it is reasonable to consider that the shaded portions of 
$\sigma(\omega)$ in Fig.~\ref{f.CTback} correspond to 
the charge-transfer type transitions as well as the other interband 
transitions with much higher energies, on the analogy of the 
band-calculation in \yto\@.\cite{bouarab1997} 
Accordingly, the remaining white portions correspond solely to 
the intraband transition within the V 3$d$ band.

In order to focus on the spectral weight of the optical conductivity 
arising from intra-3$d$-band transitions, 
we have subtracted the shaded portion in the middle of Fig.~\ref{f.CTback} 
as backgrounds, assuming an appropriate function of 
$(\omega-\Delta)^{3/2}$, where $\Delta$ has been obtained 
by fitting the lower energy tail of the O~2$p$ band in 
photoemission spectroscopy spectra of \csvo\ single 
crystals.\cite{inoue1997}

\begin{figure}
    \centerline{\psfig{file=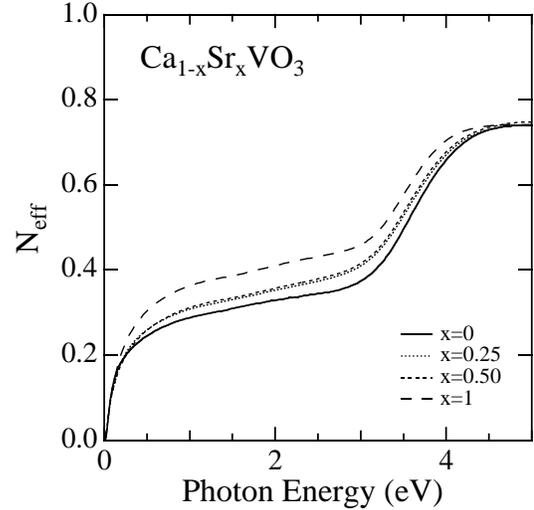,width=7cm}}
    \caption{Effective electron number per vanadium atom 
    $N_{\rm eff}$ obtained after subtracting the higher-energy 
    background (Fig.~\protect\ref{f.CTback}).}
    \label{f.Neff}
\end{figure}

A quantitative measure of the spectral weight 
has been obtained by deducing the effective electron number per vanadium 
ion defined by the following relation 
\[N_{\rm eff}(\omega)\equiv\frac{2mV}{\pi e^{2}}
\int_{0}^{\omega}\sigma(\omega^{\prime})d\omega^{\prime},\]
where $e$ is the bare electronic charge and $m$ is the bare band mass 
of a non interacting Bloch electron in the conduction band.
$V$ is the cell volume for one formula unit (one V atom in this 
system)\@.
The significance of \neff\ will be appreciated by 
considering the sum rule of the conductivity
\[\int_{0}^{\infty} \sigma(\omega)d \omega= 
\frac{\pi Ne^{2}}{2mV}\] 
where $N\equiv N_{\rm eff}(\infty)$ corresponds to the total number 
of electrons in the unit formula.
That is, \neff($\omega$) is proportional to the number of electrons 
involved in the optical excitations up to $\omega$\@.
In Fig.~\ref{f.Neff}, we show \neff\ of the \csvo\ system, 
after subtracting the higher-energy background.
Since, in Fig~\ref{f.Neff}, we have assumed $m=m_{0}$, where $m_{0}$ is 
the bare electronic mass, the total number 
$N\equiv N_{\rm{eff}}(\omega=\infty)\simeq 
N_{\rm eff}(\omega=5~{\rm eV})$ 
results smaller than 1, reflecting the difference between $m$ and 
$m_{0}$ ($m>m_{0}$)\@.
If we rather use the value $m$ obtained from LDA, 
$m\sim1.5m_{0}$ for the V~3$d$ band, we find 
$N_{\rm eff}(5~{\rm eV}) \approx 1$.
Thus, we conclude that the assumed background (shaded area in 
Fig.~\ref{f.CTback}) is reasonable to deduce the intrinsic 
contributions of the interband transition.

The initial steep rise of \neff\ is due to the Drude-like contributions 
extending from $\omega=0$.
The Drude-like contribution can be distinguished below 
$\sim$\,1.5~eV, where \neff\ exhibits a flat region.
Therefore, \neff\ at $\sim$\,1.5~eV is considered to be a good measure 
for the effective mass of the carries.
The values of $m^{\ast}/m_{0}$ estimated from \neff($\omega=1.5$~eV) 
are 3.1(6), 3.0(5), 3.0, 2.7 for $x$ = 0, 0.25, 0.50, 1, which are 
almost equivalent to the values of $m^{\ast}/m_0$ 
discussed in Sec.~\ref{subsec:mass}\@.

\begin{figure}
    \centerline{\psfig{file=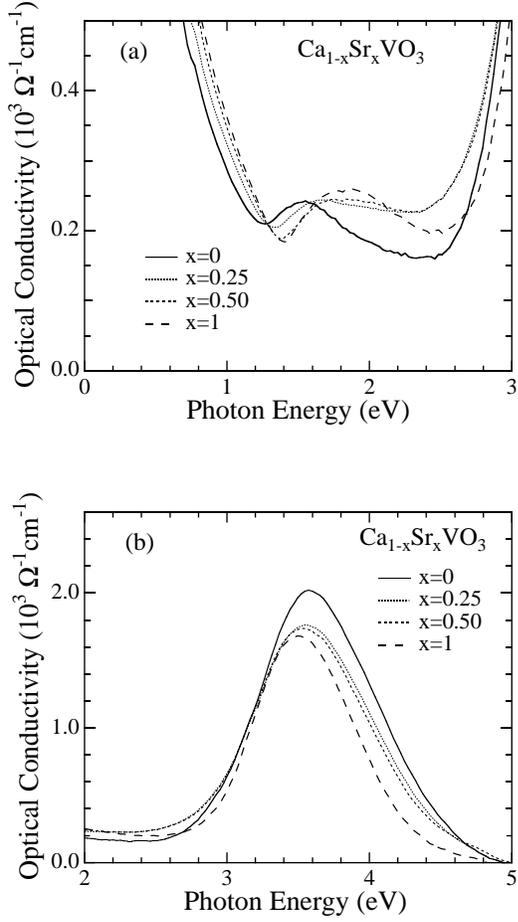,width=7cm}}
    \caption{Optical conductivity spectra of the 
    Ca$_{1-x}$Sr$_{x}$VO$_{3}$ 
    in the photon energy range of 0$\sim$5~eV. 
    The high-energy background corresponding to the interband 
    transition is subtracted. (a) a small peak at $\sim$\,1.7~eV 
    denoted as peak ``A'' in the text. 
    (b) a large peak at $\sim$\,3.5~eV denoted as peak ``B'' in the text.}
    \label{f.PeakAB}
\end{figure}

\begin{figure}[b]
    \centerline{\psfig{file=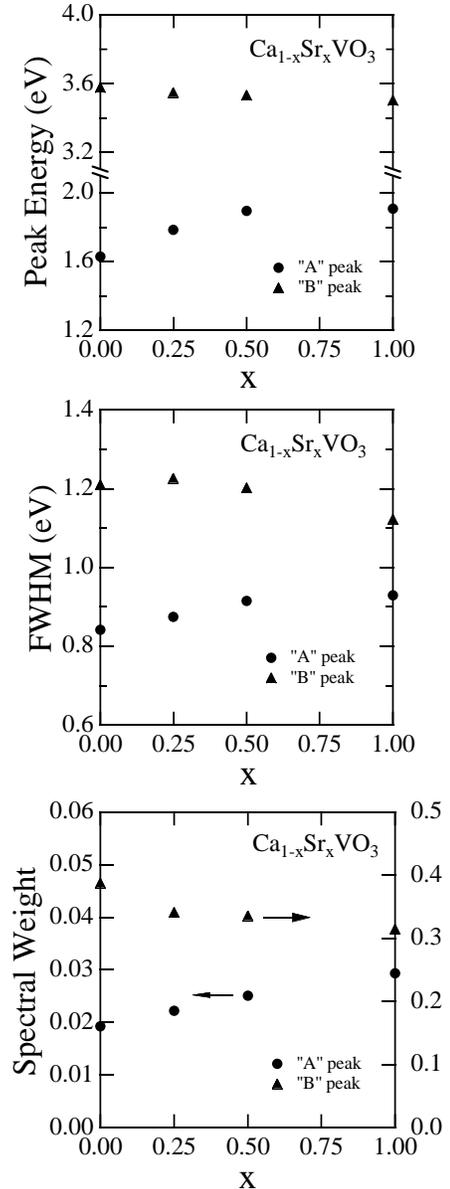,width=6cm}}
    \caption{Excitation energy, full-width at half maximum (FWHM), 
    and spectral weight of the peak ``A'' and the peak ``B'' plotted 
    as functions of $x$.}
    \label{f.ABshift}
\end{figure}

Fig.~\ref{f.PeakAB} shows the optical conductivity spectra 
$\sigma(\omega)$ of the \csvo\ single crystals ($x$=0, 0.25, 0.50, 1) 
in the photon energy range of 0$\sim$5~eV\@.
The high-energy background corresponding to the interband 
transition is subtracted.
As discussed above, $\sigma(\omega)$ ($0\leq\omega\leq5 \rm{eV}$) 
reflects only the intraband transition of the V 3$d$ electrons.
In the spectrum, there is a small feature at $\sim$\,1.7~eV, 
which we call a peak ``A'' [Fig.~\ref{f.PeakAB}(a)] 
and also a large feature at $\sim$\,3.5~eV, 
which we call a peak ``B'' [Fig.~\ref{f.PeakAB}(b)]\@.
With the increase of $x$, the excitation energy of the peak ``B'' 
shifts slightly to lower energy, 
and its spectral weight decreases; 
whereas, the excitation energy of the peak ``A'' 
shifts to higher energy, and its spectral weight increases.
In addition, the width of the peak ``A'' broadens with the increase of $x$.
Fig.~\ref{f.ABshift} shows the excitation energy, 
the full-width at half maximum (FWHM), and the 
spectral weight of the peak ``A'' and the peak ``B'' 
as functions of $x$.\cite{peak}

In the valence band photoemission spectra of the \csvo\ system, 
two features have been observed: one is a peak at $\sim$\,1.5~eV 
below $E_{F}$ and the other is the emission from a broad quasiparticle band 
which lies on $E_{F}$\@.\cite{inoue1995,morikawa1995}
The former is assigned to an incoherent emission associated with 
the formation of the lower Hubbard band 
and the latter corresponds to a renormalized 3$d$ band at 
$E_{F}$\@.
Inoue {\it et al\/}.\ reported that, upon increasing the strength 
of $U/W$ in \csvo\ system, the spectral weight is systematically transferred 
from the quasiparticle band to the incoherent part.\cite{inoue1995}
In the inverse-photoemission spectra of \cvo\ and \svo, 
Morikawa {\it et al\/}.\ have found a prominent peak at 2.5$\sim$3~eV 
above $E_{F}$ and a shoulder within around 1~eV 
of $E_{F}$\@.\cite{morikawa1995}
These features have been also assigned to the incoherent and coherent parts 
of the spectral function of the V~3$d$ electron.

These results lead us into consideration that 
the two features (the peak ``A'' and ``B'') observed in the optical 
conductivity spectra should originate in possible combinations 
of the transitions among the incoherent and coherent features of 
V~3$d$ electron around the Fermi level.

The experimental results of the optical conductivity can be compared to 
the theoretical prediction obtained by the self-consistent local-impurity
approximation of the infinite-dimension Hubbard 
model.\cite{georges1996}
The theory seems to give us a clue to understand the origin of 
the two features: the peak ``A'' and the peak ``B''\@.
According to the prediction, the optical response 
is composed of basically three contributions except the Drude part:\ 
a broad part centered at a frequency $\omega=U$\@, 
a few narrow features near $\omega=U/2$, and 
an ``anomalous'' part that is present in the range $\omega=0$ to 1~eV
approximately. 
The contribution at $U$ corresponds to direct excitations between 
the Hubbard bands,
the features at $U/2$ corresponds to excitations from the LHB to the 
empty part of the quasiparticle band and from the filled part of the
quasiparticle band to the UHB, and finally, the ``anomalous'' part 
corresponds to excitations from the filled to the empty part of the 
quasiparticle band.
In our previous paper,\cite{rozenberg1996} we analyzed the spectrum 
of \cvo\ in the light of these predictions.
The parameters $U$ and $W$, which were used in the model calculation,
were taken from 
the results of photoemission spectroscopy\cite{rozenberg1996}.
Although it was expected that the parameter $W$ would systematically 
change with composition, 
we chose to vary $U$ for the sake of simplicity, given that fits of 
equivalent quality could be obtained for the photoemission spectra.
In \cvo, which has the narrowest 3$d$ band in the \csvo\ system, 
the peaks ``A'' and ``B'' have been well described by the features 
at $U/2$ and $U$, respectively, so that the infinite-dimension Hubbard 
model seems to reproduce the experimentally obtained optical conductivity 
reasonably well.\cite{rozenberg1996}

We find that our new experimental results as summarized in 
Fig.~\ref{f.ABshift}:\ 
with the increase of $x$, 
the peak energy of the peak ``B'' shifts slightly to lower energy 
side and its spectral weight gradually increases.
The spectral weight of the peak ``A'' increases with $x$.
The peak ``A'', however, shifts slightly to higher energy. 
We shall now like to emphasize an important point. 
Our present systematic study of the \csvo\ compound gives conclusive 
evidence that we can tune the band-width of the system by controlling 
$x$\@.
The position of the peak ``B'' gives a direct measure of the value 
of $U$, and the fact that it remains almost a constant is a clear 
evidence that the ratio $U/W$ is controlled by a change of the 
band-width $W$\@.
This situation is in sharp contrast with our previous 
analysis\cite{rozenberg1996} based on photoemission data 
which did not  allow us to resolve which 
parameter was actually controlling the $U/W$ ratio.
A crucial ingredient that makes the study of the optical response so 
valuable for this analysis is that, unlike photoemission, 
it probes also the unoccupied part 
of the spectra, therefore, it is sensitive to the relative position of 
the Hubbard bands.

In order to gain some further insight in the qualitative behavior 
of the systematic evolution of our experimental data, we have used 
our initial estimates for $U$ and $W$ as input parameters in a 
calculation of the optical response of the Hubbard model and changed 
the value of $W$ instead of $U$\@. 
We shall consider the model within
the dynamical mean field theory which becomes exact in the limit of
large lattice connectivity (or large dimensionality)\@.
For convenience we have computed the optical response using the
iterated perturbation theory (IPT) method which allows for a simple 
evaluation of this quantity at $T=0$ and near the Mott-Hubbard 
transition.\cite{georges1992,zhang1993}

\begin{figure}
    \centerline{\psfig{file=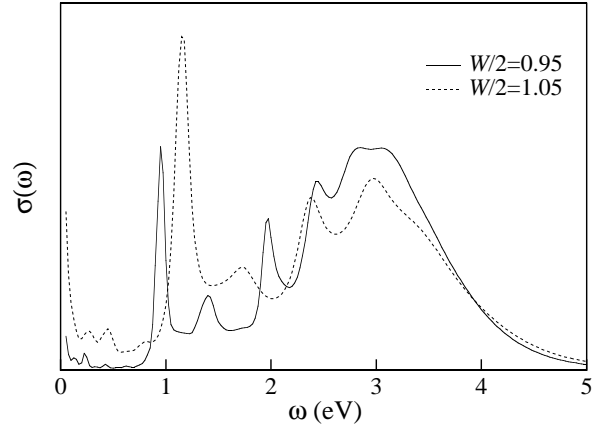,width=8cm}}
    \caption{Calculated optical 
    conductivity by IPT for the parameters $U=3$~eV and 
    $W$ indicated in the figure.}
    \label{f.largedOC}
\end{figure}

\begin{figure}
    \centerline{\psfig{file=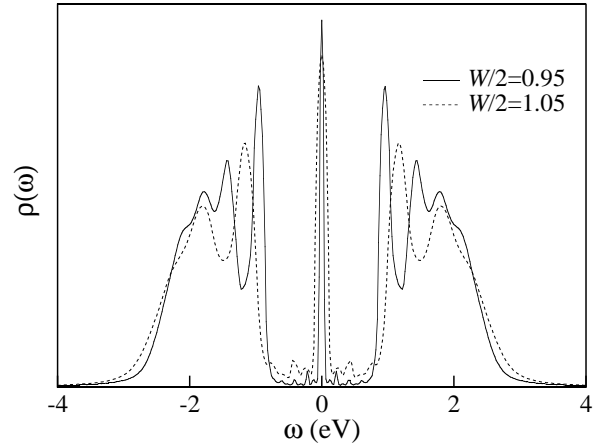,width=8cm}}
    \caption{Theoretical spectral density of states 
    obtained by IPT at $T=0$ for the parameters $U=3$~eV and 
    $W$ indicated in the figure.}
    \label{f.largedDOS}
\end{figure}

In Fig.~\ref{f.largedOC} we show the theoretical prediction using 
the value of $U=3$~eV for the
local repulsion and for the half-bandwidth 
$W/2=1.05$ and 0.95~eV for \svo\ and \cvo, respectively.
Note that the spectra do not display the Drude contribution
as it corresponds to a delta-function at $\omega=0$ since our model 
does not contain disorder and the calculation is performed at $T=0$\@.
The particular lineshape that we obtain is originated in the behavior of 
the spectral density of states that is obtained within the IPT method 
as shown in Fig.~\ref{f.largedDOS}\@.
We should point out that while 
the details of the line shape may not be correctly given by this method,
the main distribution of the spectral weight of the various contributions
and their systematic evolution are very reliably 
captured.\cite{georges1996}

We observe that the theoretical results for the systematic dependence
of the various contributions to the optical response by controlling 
the value of $W$ are in a better qualitative agreement with 
the experimental data of Fig.~\ref{f.ABshift} 
than the previous calculation where we changed the value of $U$\@.
One of the most notable improvements consists in that the unexpected 
systematic evolution of the feature at $U/2$, which shifts upward with 
increasing $W$, is well qualitatively captured.
This peculiar effect can be interpreted as a ``band repulsion'' 
between the Hubbard band and the quasiparticle band.
As we increase $W$, the latter becomes broader and ``pushes'' the 
Hubbard band further out.

However, there are still some discrepancies.
Such a large spectral weight redistribution 
is predicted to be concomitant with a large effective mass 
in the mean-field treatment of the Hubbard 
model.\cite{georges1996,brinrice}
This is, however, inconsistent with the observed effective mass 
in this system.
Moreover, in the optical conductivity spectra, 
there is apparently a notable discrepancy in respect to 
the relative spectral weight of the peak ``A'' to the peak ``B''. 
It is remarkably suppressed in the experimental data, compared 
with the theoretical data.

Other discrepancies between some experimental results 
and the prediction of the mean-field approach for the electron 
correlation were also reported 
in the photoemission spectroscopy measurements in this 
system.\cite{inoue1995,morikawa1995}
In the mean-field Fermi liquid approach, 
the renormalized quasiparticle band at $E_{F}$ should be narrowed 
with increasing the value of $U/W$,\cite{georges1996,brinrice} 
but, in those experiments, the quasiparticle band-width 
remains broad, even if the system approaches to 
the Mott transition.
Since the peak ``A'' has been assigned to the transitions associated with 
the quasiparticle band, 
the conspicuous suppression in spectral intensity of the peak ``A'' 
reflects the broadness of the quasiparticle band.
The broad quasiparticle band also accounts for the lack of the strong mass 
enhancement in this system.

As discussed in the preceding paper,\cite{inoue0000} 
the momentum-dependent self-energy becomes significant 
near the Mott transition, resulting in a reduction of the mass enhancement. 
Although our measurement cannot clarify the validity of introducing 
a momentum-dependent self-energy, 
we conclude that there must be other interactions 
not present in the mean-field treatment of the electron correlation 
in the metallic regime close to the Mott transition.

Finally, the presence of the ``anomalous'' contribution at 
low frequencies that extends down to $\omega =0$ in the theoretical 
data sheds a different light for the interpretation of 
the Drude-like response 
discussed in Sec.~\ref{subsec:mass}\@.
It may be possible to say that 
the deviation from the simple Drude model would be 
partly due to this ``anomalous'' contribution.
The origin of this effect is again traced to the presence of the incoherent
contribution coming from the low energy tails of the Hubbard bands that is 
observed in the theoretical density of states. 
However, it is experimentally very difficult to disentangle 
unambiguously the contribution of the coherent 
optical response of carriers and that of the incoherent process in 
optical conductivity, 
so this issue remains an open question.
\section{CONCLUSIONS}
This study has aimed at elucidating the electronic structure 
of the correlated metallic vanadate 
by means of the optical spectroscopy measurements.
We have synthesized the \csvo\ system to control solely 
the 3$d$ band-width without varying the band filling.

We have found that the low energy contribution to the optical 
conductivity spectra cannot be reproduced by the simple Drude model 
with the energy-independent scattering rate and effective mass.
The energy-dependent $\gamma(\omega)$ determined by the generalized 
Drude model shows relatively large energy dependence. 
However, $\gamma(\omega)$ is proportional to 
$\omega$ rather than that of the electron-electron scattering 
$\omega^{2}$\@.

The effective mass of the V 3$d$ electron has been evaluated 
from the plasma frequency.
The value of $m^{\ast}/m_{0}$ gradually increases 
with decreasing the band-width $W$\@.
However, any symptom of the critical mass enhancement has 
not been observed, even though the system is close to the Mott 
transition.

We observed two anomalous peaks in the optical conductivity spectra 
around 1.7~eV and 3.5~eV\@.
These features can be assigned to the possible combinations of transitions 
between the incoherent and coherent bands of quasiparticles 
around the Fermi level.
This large spectral weight redistribution substantiates 
the strong electron correlation in this system, which is, however, 
not concomitant with a large effective-mass enhancement.
\section*{ACKNOWLEDGEMENTS}
We would like to thank the staff of the Photon Factory for 
technical support. Thanks are also given to Y.~Nishihara and 
F.~Iga for helpful suggestions and experimental supports.

\begin{thebibliography}{99}
\bibitem{epsilon}
    In fact, we should take into account a small contribution of 
    a peak around 1.7~eV in the $\epsilon_{1}(\omega)$ spectra. 
    The peak corresponds to the ``A'' peak in the optical conductivity. 
    But here, we have disregarded the small peak.
\bibitem{peak}
    We have divided the peak ``A'' and ``B'' by fitting the 
    $\epsilon_{2}(\omega)$ data with two Gaussian. And the spectral 
    weight have defined as an amount of the contributions to 
    $N_{\rm eff}(\omega)$ from the peak ``A'' and ``B''.
\end{thebibliography}
\end{document}